\documentclass[12pt,preprint]{aastex} 
\begin{document}

\title{An ESO/VLT survey of NIR (Z$\leq $25) selected galaxies at redshifts 4.5$<$z$<$6: 
constraining the cosmic star formation rate near the reionization epoch
\altaffilmark{1}}

\altaffiltext{1}{Based on observations
obtained in service mode at the ESO VLT for the program 65.O-0445}

\author{A. Fontana, F. Poli, N. Menci}
\affil{INAF Osservatorio Astronomico di Roma, via Frascati 33,
    Monteporzio, I-00040, Italy}
\author{M. Nonino}
\affil{INAF Osservatorio Astronomico di Trieste
    via G. B. Tiepolo, 11, Trieste I-34131 Italy}
\author{E. Giallongo}
\affil{INAF Osservatorio Astronomico di Roma, via Frascati 33,
    Monteporzio, I-00040, Italy}
\author{S. Cristiani}
\affil{INAF Osservatorio Astronomico di Trieste
    via G. B. Tiepolo, 11, Trieste I-34131 Italy}
\author{S. D'Odorico}
\affil{European Southern Observatory, Karl-Schwarzschild Strasse,
    Garching bei Munchen, D-85748, Germany}

\begin{abstract}

We present the results of a VLT and HST imaging survey aimed at the
identification of $4.5<z<6$ galaxies. In the VLT data, a set of broad
and intermediate band filters has been used to select 13 high--$z$
candidates in a $Z_{AB}\leq 25$ {\it mags} catalog, over an area of about 30
arcmin$^2$. Discrimination against lower redshift interlopers (mainly
early--type galaxies at high redshift and cool Galactic stars) has been
done combining morphological and spectral classification.  This sample
has been combined with a deeper $I_{AB}\leq 27.2$  {\it mags} sample obtained from
the Hubble Deep Field campaigns.

The VLT final sample consists of 13 high--$z$ candidates, 4 of which
are identified with high confidence as $z>4.5$ galaxies. The resulting
integral surface density of the $Z_{AB}<25$ candidates at $z>4.5$ is
in the range $0.13-0.44$ arcmin$^{-2}$ and that in the highest
redshift bin $5<z\leq 6$ is between $0.07-0.13$ arcmin$^{-2}$.  In the
two HDFs, we identify at $I_{AB}\leq 27.2$ 25 galaxies in the range
$4.5\leq z < 5$ and 16 at $5\leq z\leq 6$, corresponding to surface
densities of $3.1$ arcmin$^{-2}$ and $2$ arcmin$^{-2}$, respectively.

We show that the {\it observed} $Z_{AB}<25$ UV luminosity density
appear to drop by about one order of magnitude from $z\simeq 3$ to
$z\simeq 6$.  However, if we apply a threshold to obtain an
absolute--magnitude limited sample, the UV luminosity density results
to be roughly constant up to $z\simeq 6$.

We finally show that recent semi-analytic hierarchical models for
galaxy formation, while predicting a nearly constant {\it total} UV
luminosity density up to $z\simeq 6$, under-predict the observed UV
luminosity density at $Z_{AB}\leq 25$ and over-predicts the
$I_{AB}\leq 27.2$ one. This behaviour can be understood in term of
a poor match to the slope of the UV luminosity function.

\end{abstract}


\section{Introduction}

Current findings on the production of the stellar baryon budget of the
Universe (Madau et al. 1998), have shown a complex,
gradual process, spread across a considerable fraction of the
cosmological lifetime, rather than confined into a preferred epoch in
the past.  The evolution of the rest-frame ultraviolet luminosity
density can be used to trace this process: since it is produced mainly
by O, B short-living massive stars, it is almost independent of the
star formation history of the galaxies (Madau, Pozzetti \& Dickinson
1998), but it is tied to the fraction of the current star formation
rate not extincted by dust.

Measurements of the UV luminosity density from local environment up to
$z\simeq1$ show an increase with redshift (Lilly et al. 1996, Madau
1996,  Wilson et al 2002).  At the
highest redshifts, where the results are often based on imaging surveys
and photometric redshifts, the UV luminosity density  appears to be roughly constant
between $z=1$ and $z\sim4$ ( Connolly et al 1997, Pascarelle et al
1998, Fontana et al. 1999, Steidel et al 1999, Lanzetta et al. 2002)

Here we focus upon the redshift interval $z=4.5-6$, a range which
marks a critical phase in the history of the Universe, when it was just
emerging from the reionization epoch.  Although first glances of the
early Universe have been made possible from Sloan Survey, that have
provided the first statistically defined sample of quasars extending
up to $z=6.28$, the galaxy population at these epochs is presently
nearly unexplored.

Indeed it is at these very high redshifts that it is possible to set
severe constraints on the physical mechanisms that drive the galaxy
formation and evolution. The main goal behind the study of high
redshift galaxies is the construction of a coherent picture of the
physical processes that led to galaxy evolution.  Within the framework
of gravitational instability driven by primordial fluctuation, simple
but physically motivated prescriptions have been used over the last
years to describe the main processes involved (e.g. White \& Frenk
1991; Cole et al. 1994,2000; Somerville \& Primack 1999; Menci et
al. 2002). These models are not as ``tunable'' as commonly believed
since they can not be modified freely to match the high redshift
observables without worsening the local fits (e.g. the local
luminosity function or local Tully-Fisher relation). In this respect,
any discrepancy between the model predictions and the observed
properties reveals the lack or incorrect treatment of some fundamental
processes.  In this context, it is important to perform these
comparisons directly at the highest redshifts, where the physical
processes leading to present-day galaxies are caught in act.

Despite the obvious interest in the field, and the exciting results
obtained at $z\simeq 3$ (Steidel et al. 1996), the discovery of
galaxies at $z>4.5$ has been so far serendipitous.  There are obvious
difficulties to tackle: first, the objects become progressively
fainter and rarer. Second, the multicolor ``drop--out'' criteria used
to select high--$z$ galaxies must be shifted from the UV-visual into
the near-IR bands to follow the rest--frame UV.  For the same reasons,
the spectroscopic follow--up is progressively harder. Besides this,
the number of possible interlopers increases, with early--type galaxies
and late--type stars progressively entering in the selection criteria. The
current statistics of high--$z$ galaxies reflects these problems.  Few
serendipitous objects have been identified (e.g. Spinrad et al 1998),
mostly because of a large EW emission line identified as Ly$\alpha$,
(e.g. Chen et al 1999; Hu et al. 1998, 1999, 2002). Some
identification have been later disputed on the basis of deep imaging
shortward of the (presumed) Lyman Limit (Chen et al 2000, Stern et al
2000).  In general, the search of $z\geq5$ emission line galaxies has
been shown to suffer from the strong contamination by OII emitters at
$z\simeq 1.4$ (Stern et al 2000).  Thus, the conclusions that can be drawn
from the few objects observed so far are that {\it a)} spectroscopy is
by itself not conclusive at these redshifts to derive a firm estimate
of the average cosmic star formation rate, firstly because it mostly
requires the existence of a strong Ly$\alpha$ emission line, and
because of the contamination from OII emitters; and {\it b)} deep
imaging observations are required in any case not only to select the
objects but also to validate the spectroscopic identifications.

Previous color selections at $z>4.5$ based on the photometric redshift
technique (Fontana et al. 1999, 2002) took advantage of the very deep
HST North and South samples, discovering 4 $z>5.5$ candidates in each
field down to $I_{AB}<27.5$. Evidence has come up for a comoving UV
luminosity density at $z\simeq 5$ lower by a factor of 5 than at
$z\simeq3$.  However, this estimate is based on a sample of very faint
sources selected in a small area and could be strongly affected
by the presence of large-scale structure in the ``pencil'' beams.

For this reason, we have started a relatively deep survey of galaxies
selected in the Z band to search for galaxies at $4.5 \leq z \leq 6.2$,
covering two well known fields where broad band multicolor imaging is
available, namely the extended ESO Imaging Survey (EIS) Hubble Deep
Field South and the NTT Deep Field.  We have designed a strategy based
on a combination of intermediate and broad band filters to identify
high--$z$ galaxies against the increasingly large number of interlopers.  We
emphasize that this is not a ``pre--selection'' survey, but it is
designed to provide by itself a reliable identification for the bulk
of the galaxy population at $z\geq 5$.  To increase the accuracy of
the photometric estimate of the redshift and to minimize the presence
of interlopers (mainly intermediate redshift early--type galaxies and
late type stars) in our high $z$ sample, we have included a set of
intermediate band filters namely IB691, IB834 and IB915, available at
the ESO/VLT FORS imager. These filters, although of intermediate
width, are relatively efficient since they sample spectral regions
devoid of strong sky emission lines. The adopted filter set is thus
tailored to trace the peculiar spectral features of the extremely high
redshift galaxies, i.e. the flat rest--frame UV continuum that even at
$z\sim 6$ can be sampled by the Z-IB915 color and the very abrupt
change of the shape due to the hydrogen absorption spectral breaks by
the intergalactic and interstellar medium sampled by the bluer bands.

In the following, we will adopt a $\Lambda$CDM cosmology with
$\Omega_\Lambda=0.7$, $\Omega=1$ and $H_0=70$ km/s/Mpc. All magnitudes
will be given in the AB system.

\section{Photometric database and the selection of the sample}

The dataset that we have analyzed consists of a composite sample of
optical and infrared images centered on two public deep fields: a 16.3
$arcmin^2$ area around the ESO/NTT deep field (NDF, Arnouts et
al. 1999) and a 13.6 $arcmin^2$ area around the Hubble Deep field
South (HDFS). The new data presented here have been obtained with the
spectroimagers FORS1 and FORS2, at VLT-Antu and VLT-Keyun
respectively, in the R, IB691, I, IB834, Gunn $z$ and IB915 bands, and
coupled with available UBV and near--IR images obtained at NTT
(Saracco et al 1999, (da Costa et al 1998, Fontana et al
2000). Details of the observations are given in Table 1.

Data reduction has followed the usual steps for deep imaging surveys
in dithered mode, as described for instance in Fontana et al (2000).
Since fringing effects have been found to be significant in the
reddest bands, we have applied a special care in masking even the
faintest objects prior to fringing removal. To account for the
remaining sky residuals and for the pixel--to--pixel correlation
induced by dithering, we have adopted an estimation of the noise from
sky variance inside a 3'' diameter aperture as described in Hu et
al. (1999).
 
In the overall 29.9 $arcmin^2$ area common to the optical and IR
observations, a photometric multicolor catalog of objects detected in
the Gunn $z$ band has been extracted following a recipe tailored to
obtain accurate color photometry from ground--based images with
different seeing conditions that we developed and tested on the
imaging data of the K20 survey (Cimatti et al 2002).  Total magnitudes
and colors have been estimated by the SExtractor code (Bertin \&
Arnouts 1996). For relatively bright objects, both the total flux in
the Gunn $z$ images and colors have been estimated from the
MAG\_AUTO magnitude.
For faint objects, we have used aperture magnitudes obtained by
measuring fluxes within three increasing aperture diameters of 1'',
2'' or 3'' depending on the relative distance of the nearest source.
In this case, an average correction (estimated on brighter galaxies)
has been applied to account for seeing differences among different
bands.  The choice of apertures of 1'', 2'' and 3'' has been taken on
the basis of the results obtained within the K20 dataset, where we
found that using apertures as large as 3'' (for isolated objects)
still leads to an improvement in the accuracy of photometric
redshifts. In the case of the present data, where the sample is deeper
and the seeing of the the detection image is better than the K20 $K$
images, we verified that the results do not depend on the exact choice
of the largest aperture. Indeed, we still select the same sample of $z\geq
4.5$ candidates and obtain similar estimates of the UV luminosity
densities even using catalogs based on 2'' apertures, although the
total magnitudes of individual objects may vary up to about 0.1 mags.
We limit in the following the discussion to the $Z_{AB}<25$
subsamples.

The selection of high--$z$ candidates has been performed using the
complete multiband catalogs. The spectral features that identify high
redshift galaxies are the abrupt spectral breaks due
to intervening and intrinsic HI absorption and the flat rest--frame UV
continuum long-ward of Lyman $\alpha$. In particular, the HI absorption 
produces flux dropouts in the V, R and even I bands as the redshift increases
from $z=4.5$ to $z=6$.
The selection of high--$z$ candidates is usually done
with color selection criteria sensitive to this pattern.  In our case,
due to the large number of filters adopted and to the relatively wide
redshift range covered, we have directly applied our photometric
redshift code, that is described in details in Giallongo et al (1998)
and Fontana et al (2000). The only difference respect to these papers is
that we have used here a set of spectral templates obtained with the
Pegase 2.0 code (Fioc and Rocca-Volmerange 1997), that we have found
to provide accurate redshift estimation on ground--based data sets
(Cimatti et al 2002, Fontana et al 2002). As shown in Fontana et al
(1999) and Fontana et al (2000), the selection of high--$z$ candidates
with photometric redshifts is at least as efficient as the simple
``drop--out'' technique at $z\simeq 3$, and more efficient at higher
$z$.

Potential contaminants of the selected subsample of $z_{fit}>4.5$
candidates might come either by early--type galaxies at $z>1$ (that may
mimic the large spectral breaks in the optical bands) or by cool
galactic stars that could mimic the main spectral features of galaxies
in this redshift range, as shown by Fontana et al (2000).

The contaminations from the first class of interlopers is easy to
treat with our filter set, since early--type galaxies have very bright
detections in the infrared J and K bands, so that they are
automatically excluded by our photometric redshift code.

The cleaning of the sample from the second kind of interlopers has
been achieved following a double selection criterion, based both on
morphological and spectral informations. Morphology has been taken
into account using the SExtractor {\tt CLASS\_STAR} classification
parameter (Bertin 1996), that sharply classifies the objects in our
samples down to $Z_{AB}\le24$.  In addition, the stellar spectral
library by Pickles (1998) has been used to compute the expected
stellar colors in our filter sets, to provide additional spectroscopic
criteria for the star--galaxy separation.

We have first obtained the exclusion of obvious bright stars from the
sample removing the  objects at $Z_{AB}\le24$ with {\tt CLASS\_STAR}
$\geq0.85$: the objects removed by this criteria are also better fitted
by stellar templates.  We provide a check of the reliability of this
morphology-based classification in figure 1, where the best fit
galactic spectrum is shown along with a best fit outcome from a
library of stellar spectra for four out of the whole sample of {\tt
CLASS\_STAR} $>0.85$ and $Z_{AB}\le24$ high redshift candidates.
After this selection we found no high redshift candidate at this
relatively bright magnitude cut.

In the fainter magnitude interval $24<Z_{AB}\le25$, 16 objects have
been detected by our photometric redshift code as having $z>4.5$.
Here an automatic classification becomes less reliable, since
morphology cannot be accurately estimated at these faint levels. These
candidates have been therefore inspected visually, and morphological
information has been complemented with a comparison between the best
fit $\chi^2$ of the galactic template and the $\chi^{2}_{STAR}$ of the
stellar spectra.

From this analysis, we have singled out a ``minimal'' subsample of 4
sources, that represent the most robust candidates showing clear
extended morphological features, self--consistently detected in more
than one band. On the other hand, 3 point like sources have been
excluded as having stellar full widths at half maximum in more than
one band. For these objects, the spectral energy distribution is also
more consistent with that predicted by the stellar library,
i.e. $\chi^{2}_{STAR}<\chi^{2}$.  The remaining 9 objects constitute
the sample of candidates showing ambiguous characteristics, i.e.
$\chi^{2}_{STAR}>\chi^{2}$ but compact morphology or extended
morphology but $\chi^{2}_{STAR}<\chi^{2}$.

The spectral energy distribution of 12 out of our 13 candidates are
shown for illustrative purpose in Fig.2, together with their best fit
galaxy templates.  The crucial role played by intermediate band
filters is evident especially in those cases where only upper limits
or strongly uncertain values are available for the NIR J,K bands. In
this way it is possible to sample the main features of this spectra,
i.e. the strong drop short-ward the Lyman series absorption and the
quite flat behavior long-ward the same wavelength with an accuracy
sufficient to discriminate between galaxies at $4.5<z<5$ and $5<z<6$.

In the following we will also use the photometric redshift sample of
the WFPC2--HDFN and WFPC2--HDFS fields, presented in Fontana et al
(2000) and Fontana et al (2003), to select $z>5$ galaxies at a deeper
threshold of $I_{AB}\leq27.2$. For the HDFS, the sample has been
obtained from a new optical--IR catalog that uses ultra-deep IR
observations obtained with VLT--ISAAC. The catalog extraction
procedure is described in Vanzella et al (2001), while the final
optical--IR photometric redshift catalogs are presented in Fontana et
al (2003). In both cases, we will adopt a $I_{AB}\leq27.2$ selection
criterion that ensures a small level of systematics in the estimated
magnitudes (Vanzella et al 2001).  Combining the two HDFs samples, we
identify at $I_{AB}\leq 27.2$ 25 galaxies in the range $4.5\leq z < 5$
and 16 at $5\leq z\leq 6$. A large variance still exists between the
statistics in the two field, with 17 (4) objects at $4.5\leq z < 5$
($z>5$) in the HDFS, against 8 (12) in the HDFN.

 A close up of the HDFS $z>5$ candidates is shown in Fontana et al
(2003).

We note that the new WFPC2--HDFS catalog leads to major changes
respect to the one used in Fontana et al (1999), that was based on the
Stony--Brook catalog (Lanzetta et al 2002).  In particular, three
clear candidates at $z\geq 5.5$ that were absent in the previous
catalog have been found at $I_{AB}\leq 27.2$. We found that this comes
from a systematic underestimate of the magnitudes at the faint levels
in the Stony--Brook catalog, that prevented these objects from being
included in the $I_{AB}$ selected catalog.

The resulting integral surface density of the $Z_{AB}<25$ candidates
at $z>4.5$ is in the range $0.13-0.44$ arcmin$^{-2}$ and that in the
highest redshift bin $5<z\leq 6$ is between $0.07-0.13$ arcmin$^{-2}$.
It should be emphasized that the stellar contamination of candidates
at $z>4.5$ ranges from a minimum of about 20\% to a maximum of
75\%. Thus any estimate of the surface density of very high $z$
protogalaxies should include a careful analysis of the contamination
by the faint stellar population present in a given survey.  In the two
HDFs, the resulting surface densities at $I_{AB}\leq 27.2$ is of
$3.1$ arcmin$^{-2}$ and $2$ arcmin$^{-2}$, respectively.

Our lower and upper limits on the observed number densities at
$Z_{AB}<25$ and $5<z\leq 6$ are 2--4 times higher than those predicted
in the redshift range $5.5<z\leq 6.5$ by Yan et al (2002), who assumed
that the shape of luminosity function does not change from $\simeq 3$
to $z\simeq 6$. The slightly higher value that we find is manly due to
the lower normalization adopted by Yan et al (2002), who chose to
adopt $1.37$ galaxies arcmin$^{-2}$ at $I_{AB}\leq 27.2$, while we
detect $2$ galaxies arcmin$^{-2}$, and to the slightly lower redshift
bin that we select. We also note that the ratio betwwen the number
densities measured at $Z_{AB}\leq 25 $ and at $I_{AB}\leq 27.2$ is
within a factor of two of that predicted by Yan et al (2002), which
may suggest that the slope of the UV luminosity function remains
indeed unchanged from $z\simeq 3$ to $z\simeq 6$.

\section{Evaluating the observed and predicted UV luminosity density}

We present in this section the UV luminosity density resulting from
our samples, using the usual $1/V_{max}$ formalism as described in
Poli et al (2001).  Given the good agreement between data of the two
fields, we compute the average UV luminosity density (LD hereafter) of
the total sample including both NDF and HDFS objects, in the two redshift
bins $4.5<z\leq5$ and $5<z<6$.

In both bins, the uncertainties due to possible stellar contamination
are bracketed by the two triangles in the same redshift bin. The
upward triangles represent the estimate derived from our minimal
sample. The downward ones are derived from the total sample. Error
bars are simply computed from Poisson statistics on the total number
of galaxies in each bin following the recipe in Geherels (1986) in the
case of small numbers.

Although we are focusing on the two highest redshift bins, we have
also computed the luminosity density at $2.5<z<4.5$, to show its
evolution on a complete $Z<25$ sample.  Again, photometric redshifts are
estimated from the multiband observations, and stars
have been excluded with a morphological and spectral classification
similar to the one described above.

The {\it observed} quantities are shown in Fig 3a.  As in Fontana et
al. (1999), and at variance with other works, we have not corrected the
observed values of the UV LD for incompleteness or extinction, but we
explicitly show the differential effects of the inclusion of a
magnitude limit and different dust extinction curves on the
theoretical expectations that we will discuss below.

To complement these observations, we present in Fig. 3b the same
quantities derived in the WFPC2 HDFN and HDFS $I_{AB}\leq 27.2$
samples.

At face value, the resulting picture is that the {\it observed} UV LD at
$Z_{AB}<25$ remains about constant in the redshift interval
$2.5<z<4.5$ to values of the order of $10^{26}$ erg s$^{-1}$ Hz$^{-1}$
Mpc$^{-3}$ and then decreases by a factor of 10 reaching $10^{25}$ 
(average between our minimal and total sample of $5<z<6$ candidates).

However, the comparison between the UV LD values computed in this way
at different redshifts is prone to several biases. First, the $Z<25$
or $I_{AB}\leq 27.2$ selection criterion results in a progressively
brighter cutoff to the luminosity function at increasing
redshifts. 
Second, the rest--frame wavelength span by the
Gunn $z$ band ranges from 2250~\AA at $z\simeq 3$ to 1300~\AA at $z\simeq
6$, which may introduce color--dependent selection effects when
computing the UV LD in the lowest redshift bins.  The correction of
both effects is uncertain, since it depends on the (largely unknown)
shape of the luminosity function and color properties of the high--$z$
galaxies.

In order to deal with these systematics in a clean way, we
reproduce them in the theoretical predictions as we will describe
below. A coarse correction of the varying luminosity cutoff is
possible if we compute the UV LD limited at the faintest absolute
magnitude of the highest $z=5-6$ sample, i.e. at limiting fluxes of
$1.2\times10^{29}$ erg/s and $3.6\times10^{28}$ erg/s (corresponding
to $M_{1400}\leq -19.8, -21.1$) in the $Z$ and $I$ selected samples,
respectively. The corresponding values are shown with empty symbols in
both figures. The resulting picture is that the UV luminosity density
of the brightest galaxies results to be roughly constant from $z\simeq
3$ to $z\simeq 6$.

The values of the UV luminosity densities are given in Tab.2.

\section{Discussion}

In this work we have presented the results of a pilot survey aimed at
detecting $z>4.5$ galaxies with deep multicolor images. The key
features of our approach are {\it a)} the use of an extended set of
broad (UBVRIZJK) and intermediate band ($\Delta \lambda \simeq
400$~\AA) filters centered at 6900 and 8340~\AA, {\it b)} the adoption
of conservative thresholds on the S/N for object detection and {\it
c)} the application morphological and spectral criteria: all these
aspects improve the estimates of the redshifts, and help exclude or
minimize the number of lower $z$ interlopers. When applied to the
present data set, these criteria make it possible to reject all the
brightest interlopers, that would dominate the LD, and to select a
``minimal'' sample of 4 high--confidence candidates at $z>4.5$, and a
sample of 9 additional candidates where the stellar contamination is
uncertain.  This ambiguity depends on the relative depth of our data
set: it is well within the possibilities of red--enhanced imagers at
8--m class telescope or of ACS to extend this kind of analysis 1-2
magnitudes fainter, so that the stellar contamination to the LD can be
lowered by a large amount, as in the case of deeper but smaller
WFPC2-HDF data that we present here.

Even with this ambiguity, the two selected datasets
constrain the $z>4.5$ UV LD with sufficient accuracy to show that the
{\it observed} $Z_{AB}<25$ UV LD drops by about one order of magnitude
from $z\simeq 3$ to $z\simeq 6$.  This drop is largely due to the
progressively brighter cutoff in the rest frame luminosity function:
if we correct for this incompleteness, the UV LD appears to be roughly
constant from $z \simeq 2.5$ up to $z\simeq 6$.

The present results are apparently in contrast with recent findings by
Lanzetta et al (2002), who make use of the same WFPC2--HDF data and
apply photometric redshifts, that claim the global SFR to steadily
increase up to $z=12$.  Unfortunately, the overall approach and
techniques adopted are so different that it is difficult to make a
clean comparison. First, we note that the use of wide aperture
magnitude and the adoption of a much higher S/N threshold to the
catalog strongly reduce the systematics in the WFPC2 data induced by
surface brightness dimming, that otherwise require the complex and
model--dependent correction applied by Lanzetta et al (2002).
Besides, we draw our conclusions from a homogeneous sample, that
includes only objects selected down to the same absolute magnitude
limit, rather than correcting for the incomplete coverage of the
luminosity function, as done by Lanzetta et al (2002).  Finally, the
photometric redshift distribution obtained by Lanzetta et al (2002) is
markedly different from our own, and peaks at $z\simeq 0$, a factor
that may lead to an overestimate of the faint end of the SFR
distribution function that is used to correct the high--$z$ data.
Despite all these differences, it is to be noted that the two results
are still consistent when the more conservative estimate of Lanzetta
et al (2002) is compared with our data in the appropriate redshift
range $z\simeq 3$ to $z\simeq 6$: given the overall uncertainties both
analyses suggest that UV LD is roughly constant in this redshift
range.

The question that naturally arises is whether a constant UV LD up to
$z\simeq 6$ may be compatible with the hierarchical scenario of galaxy
formation. To discuss this point, we present in both panels of fig. 3
the UV LD predicted by the CDM semianalytic model described in Menci
et al. (2002). This model is based on the Cole et al (2000) recipes,
with an additional improved treatment of aggregation of satellite
galaxies in common DM haloes.  We first show that the {\it total} SFR
derived from this model (thin solid lines) is nearly constant from
$z\simeq 2$ to about $z\simeq 5$, and then fades by only a factor
about 5 at $z=6$.

To allow a fair comparison with our data, we have applied the effects
of magnitude limit cut and dust extinction directly to the theoretical
model. The effect is shown by the thick solid lines in Fig3a and
Fig3b. The shaded area show the possible effects of dust extinction,
ranging from no dust (upper lines) to SMC-like extinction curve (lower
lines). Please note that the inclusion of dust correction effects in
the theoretical model decreases the average UV LD respect to the
unextincted amount.  It appears that while the CDM model broadly
encompasses the observed values, it progressively under-predicts
the LD observed in the bright $Z_{AB}<25$ sample, while over-predicts
the LD observed in the faintest $I_{AB}<27.2$ sample. This behavior
can be understood in term of a poor match to the slope of the UV
luminosity function. At the brightest magnitudes, the CDM model
under-predicts the number of bright sources, while it over-predicts
the number of fainter sources dominating the deeper sample. This is
analogous to what already found at $z\simeq 3$ and confirms a general
trend of this version of CDM models to under-predict the amount of
star--formation rate in high redshift massive objects (Somerville and
Primack 2001, Poli et al 2001 and Menci et al 2002, Cimatti et al
2002). 

The origin of this discrepancy is likely tied to the lack or to the
oversimplified treatment of fundamental physical process. Indeed, we
remark that the basic recipes adopted in this model already concur to
enhance the SFR in high redshift massive objects, and that these
aspects are further boosted with respect to the models that we used in
Fontana et al (1999). The star formation rate is computed as $\dot
M_*=M_g/\tau_*$, where $M_g$ is the amount of cool gas and the
time-scale $\tau_*$ is proportional to the dynamical time $\tau_{dyn}$
and to the galaxy circular velocity $V_c$ as
$\tau_*=\epsilon_*^{-1}\,\tau_{dyn}\,\big(V_c/200\,{\rm
km/s}\big)^{\alpha_*}$. Since both $M_g$ and $1/\tau_{dyn}$ increase
with redshift, and $\alpha_*=-1.5$, these models naturally predict
that the star-formation rate increases with redshift (for a given
galaxy mass), and is more efficient in massive galaxies (at a given
redshift). In addition, feedback effects are also a strong function of
$V_c$, since they scale as $\big(V_c\big)^{-5.5}$, and again strongly
favor the more massive objects. These basic recipes, combined with the
effects of the biased process of galaxy formation induced by
hierarchical merging, already conspire to boost at high $z$ the SFR in
massive objects, with respect to the less massive one.

In the present context, it is not possible to flatten the high--$z$
luminosity function by simply changing the free parameters of the
model, since one rapidly worsens the fit to the local observables (Cole
et al 2000). 

Other physical processes that are important in the high--$z$ Universe
are not included in our model. On the one hand, mechanical and
ionizational feedback on the intergalactic medium by early galaxy and
QSO formation is not included in our rendition, a process that is
expected to quench the SFR in low mass objects and hereby to flatten
the low luminosity side of the LF. On the other hand, molecular
cooling is efficient at high $z$ and not included here. Another
possibility that has been proposed is that the starburst efficiency
increases during major mergers.  These ``starburst'' models are known
to increase the bright side of the LF at $z=3-4$ (Somerville and
Primack 1999), but require new additional free parameters to be
introduced in the models, and may become less efficient at $z>5$, when
major merging events are very rare. The challenge of the next years is
to include all these processes in a self consistent picture that
reproduce the high--$z$ observables.

\acknowledgments

The VLT data used in this work have been mostly obtained in the
framework of the proposal 65.O-0445: we are grateful to the VLT
operators and scientists that have performed the observations in
service mode. The Gunn~$z$ observations of the NDF have been obtained
during the Science Verification Phase of FORS2-VLT. 
UBV and IR images of the HDFS have been obtained by the EIS survey. The UBV
observations of the NDF were performed in SUSI-2 guaranteed time of
the Observatory of Rome in the framework of the ESO-Rome Observatory
agreement for this instrument.

We thank the referee, R. Windhorst, for careful reading of the
manuscript and useful comments.
\bigskip


\noindent

\newpage
\begin{table*}	 
\caption{Summary of the Observational Data}  
\label{field} 
\begin{tabular}{|l|lcc|lcc|} 
\tableline 
~ & \multicolumn{3}{c|}{NTT DF}& \multicolumn{3}{c|}{HDFS}\\
Filter & Instrument &Seeing & mag. lim.$^{(1)}$ & Instrument&  Seeing & mag. lim.$^{(1)}$\\
\tableline 
U &NTT-SUSI2 & 0.97& 27.5& NTT-SUSI2$^{(2)}$ & 1.02& 28.3 \\
B & NTT-SUSI2 & 1.26 & 28 & NTT-SUSI2$^{(2)}$ & 0.92& 28.0 \\
V & NTT-SUSI2 & 1.12& 27.6& NTT-SUSI2$^{(2)}$ & 1.04& 28.1  \\
R & VLT-FORS1 & 0.88& 28.4 &VLT-FORS1  & 0.86& 27.9 \\
IB691 & VLT-FORS1 & 0.55 & 28.1&VLT-FORS1  & 0.68& 27.8 \\
I & VLT-FORS1 & 0.81 & 27.4 &VLT-FORS1  & 0.65& 27.2\\
IB834 & VLT-FORS1 & 0.63& 27.2& VLT-FORS1  & 0.82& 26.6\\
z &VLT-FORS2$^{(2)}$ & 0.65 & 27.1& VLT-FORS1 & 0.65  &26.2 \\
IB913 & VLT-FORS2$^{(2)}$& 0.65& 26.4 & -&- & -\\
J & NTT-SOFI$^{(2)}$ & 0.77 & 25.7& NTT-SOFI$^{(2)}$ & 1.04& 25.8\\
H &- &- & - & NTT-SOFI$^{(2)}$ &  1.03& 24.43\\
K & NTT-SOFI$^{(2)}$ & 0.7& 25.1 & NTT-SOFI$^{(2)}$ & 1.06& 24.87\\

\tableline 

\end{tabular} 

(1): Computed in the  AB system, at 1$\sigma$, in 2 arcsec.  Limiting magnitudes have
been estimated from the photometric catalogs used in the paper, and
defined as the typical value at which $\Delta m = 1.08/3$. \\ (2):
Public data, from {\sf http://www.eso.org}

\end{table*} 

\newpage

\begin{table*}	 
\caption{Measured UV luminosity density at $z\geq 2.5$}  
\label{tabuv} 
\begin{tabular}{|c|cc|cc|} 
\tableline 
~ & \multicolumn{2}{c|}{VLT: $Z_{AB}\leq 25$}&\multicolumn{2}{c|}{HDFs: $I_{AB}\leq 27.2$}\\
Redshift & \multicolumn{2}{c|}{log $\rho_{1400} $(erg s$^{-1}$ Hz $^{-1}$ Mpc$^{-3}$)} &   \multicolumn{2}{c|}{log $\rho_{1400} $(erg s$^{-1}$ Hz $^{-1}$ Mpc$^{-3}$) }\\
~ & $Z_{AB}\leq 25$  & $M_{1400}\leq -21.1$ & $I_{AB}\leq 27.2$  & $M_{1400}\leq -19.8$\\
\tableline 
3 & 26.07 & 25.36 & 26.22 &26\\
4 & 25.94 & 25.5 & 25.74 & 25.47\\
4.75 & 25.52$\div$25.74$^{(1)}$ & 24.61$\div$25.31$^{(1)}$ & 25.78 &25.65\\
5.5 & 24.73$\div$25.26$^{(1)}$ & 24.73$\div$25.26$^{(1)}$ & 25.43 & 25.43 \\
\tableline 

\end{tabular} 

(1): The two numbers correspond to the minimal and whole  sample, as defined in the
text.

\end{table*} 

\newpage

\begin{figure}
\plotone{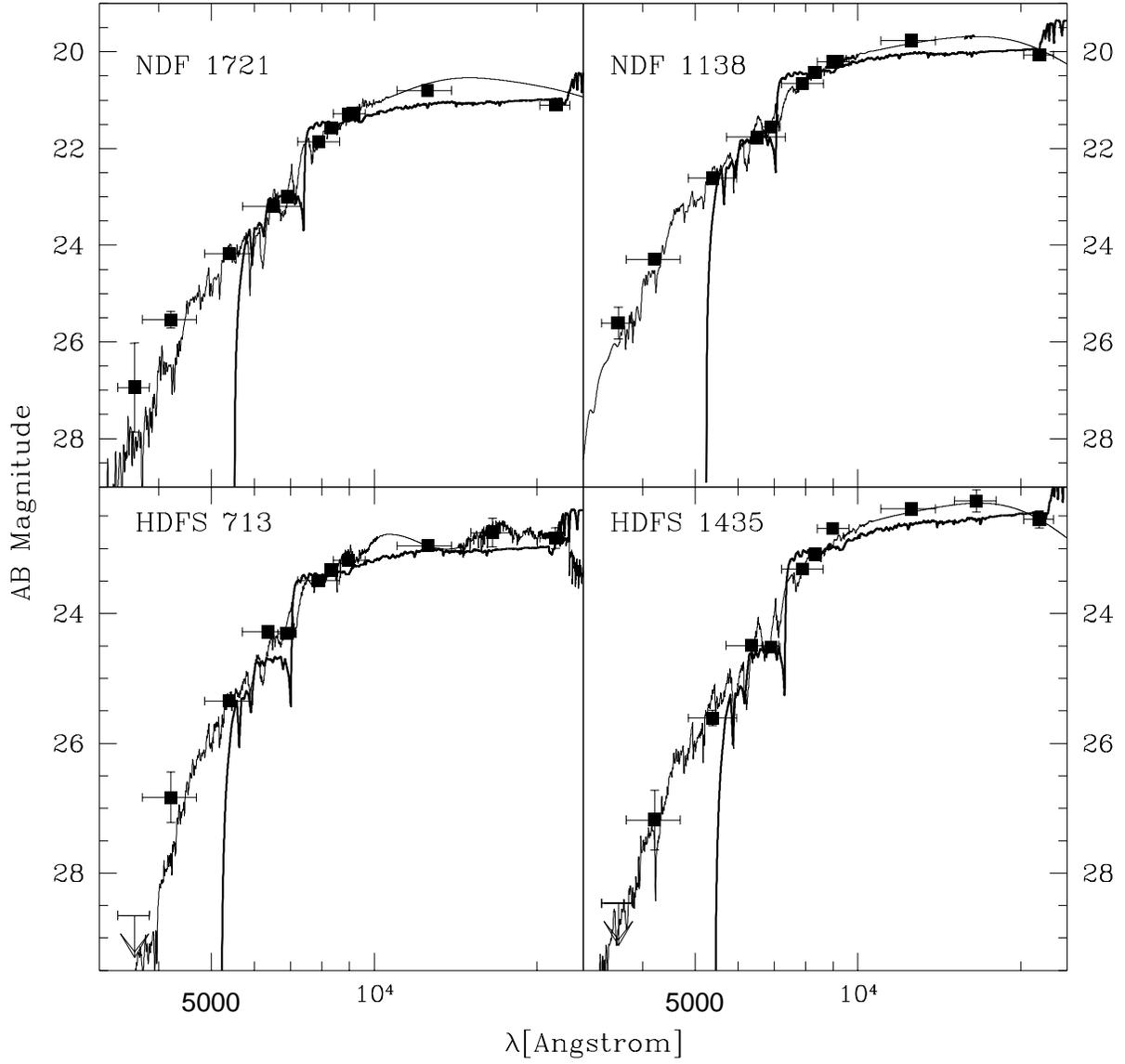}
\caption{
Best fit spectral templates from the Rocca-Volmerange library (thick line)
and from a library of stellar spectra (thin line) for a sample of 
{\tt CLASS\_STAR} $>0.85$ and $Z_{AB}\le24$ high redshift candidates. 
\label{fig1}
}
\end{figure}

\newpage

\begin{figure}
\plotone{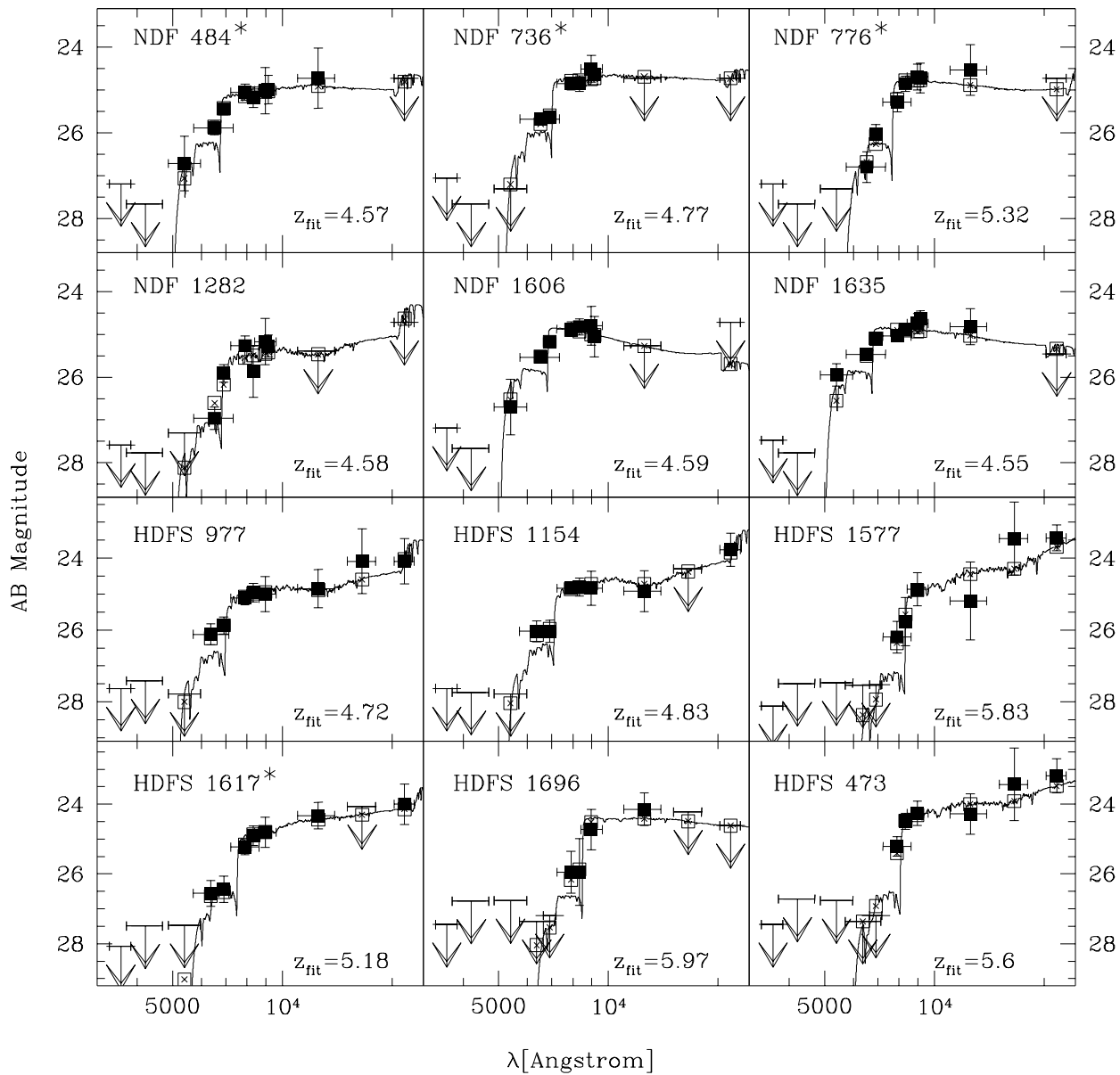}
\caption{ Best fit spectral templates for 12 out of our 13 high
redshift candidates.  Filled squares are measured AB magnitudes, 1
sigma upper limit values are displayed as downwards arrows. Empty
squares represent the mean value of the spectrum inside the
filters. Objects with asterisk belong to the ``minimal'' sample.
\label{fig2}
}
\end{figure}

\newpage

\begin{figure}
\plotone{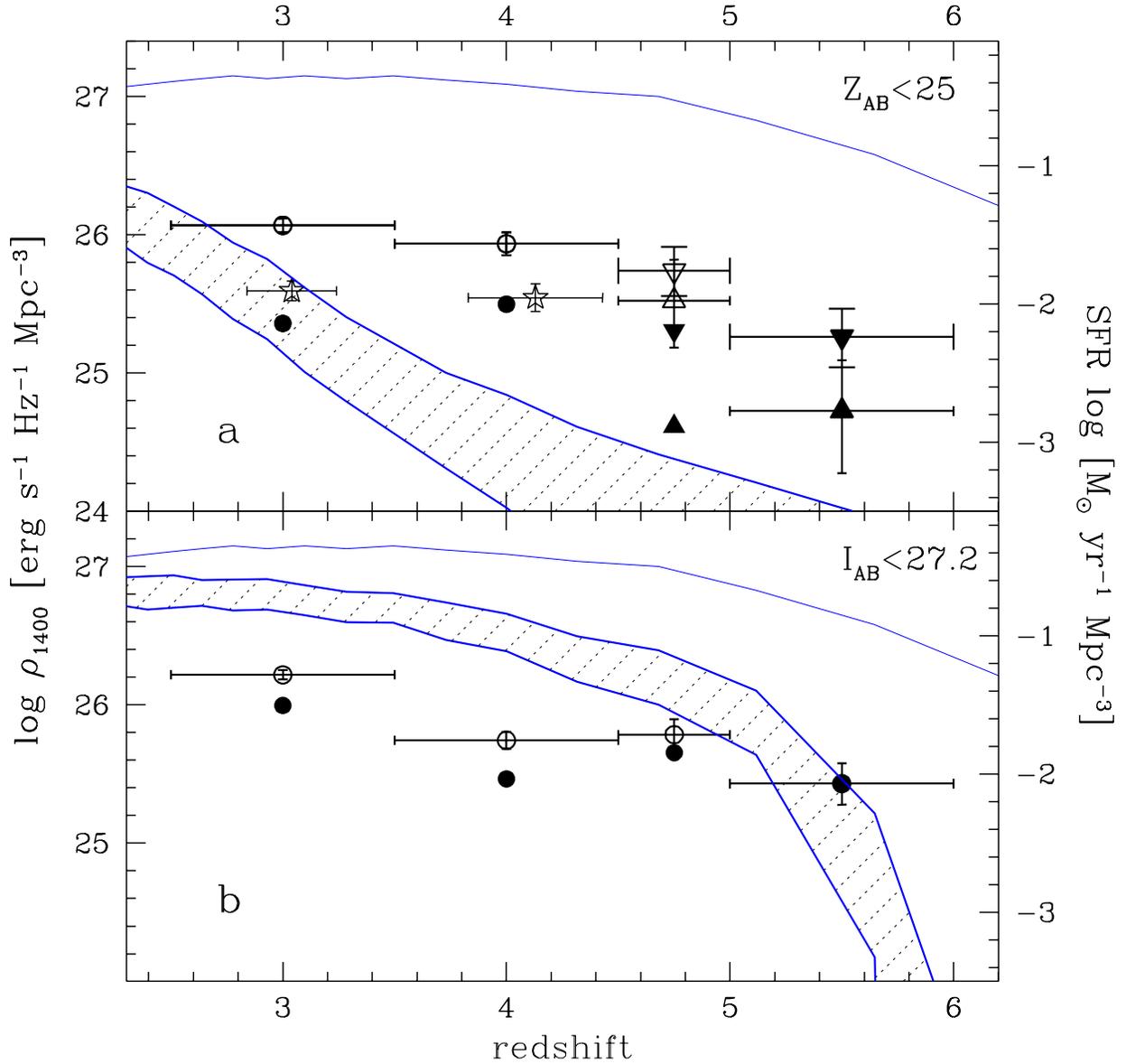}
\caption{ Observed and predicted UV luminosity densities as a function
of redshift. {\bf a)} Data from our $Z_{AB}\leq 25$ survey. Upward
triangles represent our estimates from the minimal sample, downward
triangles from the whole sample (see text for details). Empty points
represent the observed quantites. Filled symbols are the same
quantities computed in a homogeneous sample that
includes only objects down to the faintest absolute magnitude of the
highest $z=5-6$ sample (see text for details).  Empty stars represent
the values from the Steidel et al. 1999 spectroscopic sample of
galaxies at $R\leq 25$ shown for comparison.  The continuous curves
delimiting the shaded area represent the predictions at $Z_{AB}\leq
25$ by our fiducial model adopting different extinction curves for
dust absorption.  The thin curve shows the total UV (unobscured)
luminosity density associated with the total star formation rate of
the fiducial model. {\bf b)} UV LD from the WFPC2 HDF-N/S sample of
$I\leq 27.2$ galaxies. Symbol meaning and and predicted curves are as
in a)
\label{fig3}
}
\end{figure}

\end{document}